\documentclass[aps,prl,preprint,groupaddress]{revtex4}

\usepackage{bbm}
\usepackage{graphicx}
\usepackage{subfigure}
\usepackage{amsmath}

\begin{document}

\title{Long-coherence pairing of low-mass conduction electrons in copper-substituted lead apatite}

\author{Jicheng Liu$^{1}$, Chenao He$^{2}$, Yin-Hui Peng$^{3}$, Zhihao Zhen$^{3}$, Guanhua Chen$^{3}$, Jia Wang$^{1}$, Xiao-Bao Yang$^{3}$, Xianfeng Qiao$^{2,4}$\footnote{\url{msxqiao@scut.edu.cn}}, Yao Yao$^{2,3}$\footnote{\url{yaoyao2016@scut.edu.cn}}, and Dongge Ma$^{2,4}$}

\address{$^1$ School of Minerals Processing and Bioengineering, Central South University, Changsha 410083, China\\
$^2$ State Key Laboratory of Luminescent Materials and Devices, South China University of Technology, Guangzhou 510640, China\\
$^3$ Department of Physics, South China University of Technology, Guangzhou 510640, China\\
$^4$ Guangdong Provincial Key Laboratory of Luminescence from Molecular Aggregates, Guangdong-Hong Kong-Macao Joint Laboratory of Optoelectronic and Magnetic Functional Materials, South China University of Technology, Guangzhou 510640, China}

\date{\today}

\begin{abstract}
Two entangled qubits emerge as an essential resource for quantum control, which are normally quantum confined with atomic precision. It seems inhibitive that in the macroscopic scope collective qubit pairs manifest long coherence and quantum entanglement, especially at high temperature. Here, we report this exotic ensemble effect in solid-state sintering lead apatite samples with copper substitution, which have been repeatedly duplicated with superior stability and low cost. An extraordinarily low-field absorption signal of cw electron paramagnetic resonance (EPR) spectroscopy stems from low-mass conduction electrons implying the coherence of cuprate radicals can be long-termly protected. The pulsed EPR experiments exhibit triplet Rabi oscillation from paired cuprate diradicals with the coherence time exceeding 1 microsecond at 85K. We believe these appealing effects are sufficiently promising to be applied for scalable quantum control and computation.
\end{abstract}

\maketitle

Manufacture of applicable quantum computers demands for robust, reproducible and thermally stable quantum units, and long coherence time essentially matters. Macroscopic quantum effects such as superconductivity therefore attracted much attention, as superconducting circuits arose for establishing quantum computers and dramatically developed in the last several years \cite{23,24,25,26}. However, it has to work at extremely low temperature to suppress the influence of thermal environment, since the thermalization subject to complicated many-body interactions acts as a primary predator of coherence. So far, it seems impossible to realize efficient quantum control in a macroscopic bulk material at ambient circumstances due to the fragility of collective quantum coherence.

In order to keep coherence alive, alternatively, researchers prefer to quantum confine and isolate the qubits as clean as possible to weaken the interactions, with instances including ultracold atoms, ion traps, quantum dots, colour centers, molecular magnets and so on \cite{13,14,15,16,17,18,19,20,21,22}. Based upon spectroscopies of electron paramagnetic resonance (EPR) and nuclear magnetic resonance (NMR), two promising platforms to investigate radical spin dynamics, recent experiments associated with quantum computations have realized various schemes of multi-qubit gates as well as detection of effective entangled states preparation \cite{1,2,3,4,5,6,7,8,9}. Nevertheless, due to the isolation and the steep full of coherence time following scale and temperature increase \cite{10,11,12}, a great challenge turns out for large-scale fabrication, and thus a solid-state bulk material without quantum confinement that intrinsically possesses sufficiently long coherence time must be useful. Herein, we report the cuprate diradicals that are embedded in a lead apatite framework can be a perfect new candidate.

Cu-substituted lead apatite (CSLA) was recently claimed to be a room-temperature superconductor by Lee et al. \cite{lee2023roomtemperature,lee2023superconductor}. We basically followed their procedure to synthesize our CSLA samples, with $1:1$ mole ratio of Pb$_2$(SO$_4$)O and Cu$_{3}$P, whose powder XRD pattern is shown in Fig.~1(a).
The resulting characteristic peaks are well consistent with those of Pb$_3$(PO$_4$)$_2$, Pb$_4$(PO$_4$)$_2$(SO$_4$), Pb$_{4.5}$(PO$_4$)$_3$, Cu$_2$O and CuO, respectively. Fig.~1(b) shows the magnified pattern of the sample exhibiting a peak migration due to the lattice distortion caused by copper substitution. As shown in Fig.~1(c) and (d), the sample manifests multi-phase mixture, as the Energy Dispersive Spectra show the main phases might be CSLA (Spot 1 and Spot 2), mixture of copper metal and cuprate (Spot 3) and Cu$_2$O (Spot 4).

It is worth mentioning that, in the procedure reported by Lee et al. \cite{lee2023superconductor}, the mixed powders were sealed in a quartz tube with a vacuum of $10^{-5}$ torr for the sake of an oxygen-free environment to minimize the production of Cu$^{2+}$. This is essential as it is much easier for Cu$^+$ to substitute lead. The cooling procedure has however not been clearly described in the literature, and in our synthesis, the quartz tube was first taken out and cooled in liquid nitrogen after roasting at 925~$^{\rm o}$C for 15h. The XRD pattern after this step is well consistent with that reported by Lee et al. Afterward, the sample was taken out of the quartz tube and placed in an alumina crucible and roasted in a tubular furnace at 925~$^{\rm o}$C for 5h in the atmosphere, following with a final slow cooling. In this additional step, the substituted coppers are further oxidized to divalent subject to the microwave response and other possible magnetic and electric responses.

Due to the large atomic radius, the lead apatite form an ionic framework, and the doped coppers occupy some sites of leads to achieve firm structure with strong interior stress. It was stated that the cuprates constitute a quasi-one-dimensional superconducting chain \cite{10.1126/science.abf5174}. With our CSLA samples, however, we did not observe a notable superconducting Meissner effect which may be hidden in the observed paramagnetic or diamagnetic signals, and the bulk resistance is extremely high ($\sim{\rm M\Omega}$) hindering electric measurements. As the main component, the lead apatite framework does not directly contribute to the superconductivity, largely weakening the macroscopic measurements of DC or AC magnetic susceptibility and conductivity. Some reported samples have got very high conductivity \cite{hou2023observation, zhu2023order}, probably because the surface deposition of copper metals as also observed in our samples without further oxidation which do not show detectable EPR signals. In order to properly characterize the samples, therefore, we utilized EPR spectroscopy which was closely equivalent to measuring AC magnetic susceptibility by finely distinguishing different constitutions \cite{PhysRevB.36.7241, PhysRevB.36.2361}.

As shown in Fig.~\ref{fig2}(a), X-band cw EPR spectrum of the CSLA sample displays a wide peak at magnetic field from 2700 to 3400~Gauss mixed with a sharp radical response at 3350~Gauss with the Land\'{e} $g$ factor being 2.059. Through detailed analysis discussed below, Cu$^{2+}$ ions with $3d$ electrons are residing on the substituting sites and combining with oxygens on the interstitial sites, generating the inhomogeneous broadening of the peak. Another candidate of radical turns out to be the oxygen vacancy, a hole with local spin-1/2 behaving like a radical cation, which in our samples is a troublemaker for the magnetic measurements but can be useful as a standard reference for other high spin states. At exactly the position of half magnetic field, there is a strong half-field peak even at room temperature, implying the cuprate radicals are coherently paired and the absorption of microwave photons refers to triplet diradical states. This serves as the essential finding in this work, as stable and reproducible diradicals at room temperature are seldom seen in bulk materials because two radicals always prefer to form closed-shell chemical bonds.

It has also been stated that an absorption of microwave at magnetic field of 0--1000~Gauss could be solid evidence for the appearance of superconducting gap \cite{lee2023roomtemperature}. We also observe super broad low-field absorption signals at all temperatures we measured in some samples, which have to be further justified in the future. By changing the scanning rate of magnetic field, all these broad signals manifest strong hysteresis excluding contributions of paramagnetism and soft magnetism as claimed by other researchers \cite{10.1002/adfm.202308938, Wang2023, Guo2023}.

Remarkably, there is a strong Dysonian-shaped asymmetric peak at field of 406~Gauss which is much narrower than others. The effective $g$ factor is 16.949 and the relevant effective mass $m_{\rm D}$ (rest Dirac mass) is 0.12$m_0$ with $m_0$ being the mass of free electron \cite{PhysRevLett.114.186401}. In order to determine the origin of this low-mass signal, we increased the power of microwave and observed it was independently changed from asymmetric Dysonian shape to symmetric Lorentzian. The asymmetric lineshape normally stems from the dispersive component of magnetic susceptibility induced by electric-field component of microwave, so it implies the real part of AC magnetic susceptibility $\chi'$ is significantly suppressed by the increasing power. Furthermore, by increasing temperature as illustrated in Fig.~\ref{fig2}(b), the peak was found moving to even lower field and the $g$ factor was further increased, excluding contributions of ferromagnetic resonance. The ratio $A/B$ between the maximum and minimum intensities of the peak is defined as the asymmetric ratio of the Dysonian shape, which together with the linewidth are not sensitive to the temperature, meaning the radicals are spatially localized with unchanged coherent length. All these results suggest it is a conduction electron spin resonance (CESR) \cite{PhysRevB.93.155114CESR}, which in normal cases appears in good conductors with localization in nano-structures but the peaks are always located close to that of free radicals \cite{PhysRev.98.349Dyson}. To our best knowledge, this spectrum of low-mass conduction electron has been merely observed in topological insulator at low temperature \cite{PhysRevB.93.155114CESR}, in which the conduction electrons are topologically protected. Considering that in CSLA the cuprate radicals are the most possible species that do conduct, it is then reasonable that the long coherence of cuprate diradicals discussed below is also locally protected in the same manner.

Two microwave pulses with phases of $\pi/2$ and $\pi$ produce a spin echo signal, and given sufficiently long coherence time, the change of time interval of two pulses gives rise to the Rabi oscillations, which are displayed in Fig.~\ref{fig3}(a) at both 3350 and 3000~Gauss. By Fourier transformation it is clear that the oscillation at 3000~Gauss is solely dominated by one mode while that at 3350~Gauss is mixed with two modes. Rabi oscillations rely on quantum coherence which is basically fragile in bulk materials under warm environment. The detection of coherence time $T_2$ is thus shown in Fig.~\ref{fig3}(b), which manifests itself to be 1.185~$\mu$s at 85~K, and up to 260~K it is still longer than 100~ns (not shown). This timescale is sufficiently long to manipulate nearly 100 quantum gates of 12~ns $\pi/2$-pulses with moderate fidelity. By Fourier transforming this $T_2$ curve, the low-frequency modulation of the echo dynamics was determined to be from sulphur, lead and phosphor, by their individual nuclear gyromagnetic ratios. The absence of oxygen is mainly due to the rare content of $^{17}$O isotope. Once the radicals are residing on, the modulation frequency of copper would be far beyond the detection range, so the coherent echo signals can be safely regarded to stem from cuprates and oxygen vacancies. In addition, the relaxation time $T_1$ was also measured to be 2.274~$\mu$s at 85~K.

To further figure out the physical scenarios of two Rabi modes at 3350~Gauss, the Rabi frequency versus root of microwave power is displayed in Fig.~\ref{fig3}(c). It is remarkable that the slope of the high-frequency mode is closely $\sqrt{2}$ times larger than that of low-frequency one, demonstrating the former is from spin one and the latter from spin half. Considering all possible candidates of radicals, long-coherence pairing of cuprates solely matters for these triplet signals. Combined with the half-field cw signal, it implies the spatially separated cuprate diradicals are completely entangled with each other, which is highly peculiar because this quantum entanglement is not prepared in an artificial quantum confined system but observed in a realistic solid-state bulk material at relatively high temperature. Ye et al. have recently demonstrated the local pairs without global phase coherence indeed exist before the transition to superconductivity \cite{Ye2023}, which can be applied to explain the absence of macroscopic conductivity in our samples and suggest the possible route to develop. Moreover, via finely changing the pulse duration, it is able to separately excite spin one and half, and Fig.~\ref{fig3}(d) shows the resulting echo-detected field-swept spectra. The spin half signal sharply peaking at 3350~Gauss most possibly stems from oxygen vacancies. The spin one signal from cuprate diradicals is much wider covering from 2700 to 3400~Gauss, consistent with the cw EPR.

In theory, we analyzed the role of cuprates and oxygen vacancies (V$_{\rm Os}$) in CSLA, by calculating all double V$_{\rm Os}$ configurations in supercell of 1$\times$1$\times$2. There are two kinds of oxygen: One is in phosphate and the second (O2) is distributed at the edge of the lattice as shown in Fig.~\ref{fig4}(a). Two stable structures with double V$_{\rm Os}$ are marked as ABAB and AABB respectively, and the total energy of the latter is merely 36~meV lower than that of the former, suggesting that before the substitution of coppers, the two structures ABAB and AABB coexist in lead apatite in a multi-phase manner. We then calculated different copper distribution in both ABAB and AABB structures, and the energy variation with Cu-Cu distance is depicted in Fig.~\ref{fig4}(b), which reveals a notable energy advantage of Cu-substituted AABB over ABAB. For ABAB, the copper and surrounding oxygen atoms enable a local octahedral structure, and for AABB, they form a slightly distorted Cu-O plane, consistent with the fundamental unit of the stablest $P$42/$mmc$-CuO \cite{10.1063/1.4812323}. Most importantly, the stablest Cu-Cu distances in both ABAB and AABB are longer than 7~{\AA} suggesting an exotic long-range pairing of cuprates. Fig. \ref{fig4}(c) shows the electronic band and the projected density of states ($p$DOS) of these two structures. A flat band crosses the valence band maximum in ABAB, while AABB possesses a semiconducting feature with 1.22~eV band gap. Combined these results, it was realized that AABB should be the primary phase of CSLA while with an appropriate cooling process, there will also be considerable ABAB constitutions. The $p$DOS of ABAB further shows that the hybridization between copper and oxygen gives rise to a dominant copper $3d$ orbital at the Fermi energy where a level crossing appears implying the possible contribution of conduction Dirac electrons. With the transformation of the local Cu-O octahedron into Cu-O plane, the ground state of AABB is dominated by oxygen $2p$ orbitals which should mainly be responsible for the sharp radical signals.

Before ending, we would like to discuss more about the scalability of CSLA. Different from the quantum confined systems, CSLA is synthesized by common solid-state sintering approach which can be easily applied to large scales. Giving raw materials purchased from different companies our results are highly consistent and reproducible, convincing us that it can be enlarged with moderate performance. Furthermore, the lead apatite is actually an inclusive framework, and except for cuprates, other elements can also be candidate and form some kinds of internal heterojunctions, which may respond to more interesting microwave sequences. In addition, as a bulk semiconductor one can also manipulate the samples with gate voltage and light field, making more possibilities.

\section{Methods}

\textbf{Characterizations}

X-ray diffraction (XRD) measurements were carried out in an X-ray system (Rigaku SmartLab SE) with copper K$\alpha$ radiation ($\lambda=1.78897$~\AA, 35~kV and 40~mA). Scanning electron microscope (SEM) pictures were collected using field emission scanning electron microscope (TESCAN, MIR3) equipped with energy-dispersive X-ray spectrometer (EDS).

\textbf{EPR measurements}

The samples were transferred into quartz EPR tubes with an inner diameter of 2.7~mm (outer diameter of 3~mm) in glove box, and then the quartz EPR tubes were encapsulated by ultraviolet curing adhesive. The cw EPR spectroscopy was measured on an EPR spectrometer (Bruker ELEXSYS E580) operated at the X-band (9.667~GHz) and outfitted with a dielectric resonator (ER-4118X-MD5). The microwave power is 15~dB, and the modulation amplitude is 5~Gauss at 100~kHz. The magnetic field was corrected by using a BDPA standard (Bruker E3005313) with $g=2.0026$. Low temperature environment was realized by an Oxford Instruments CF935 continuous-flow cryostat using liquid nitrogen. The temperature was controlled by an Oxford ITC4 temperature controller with accuracy of $\pm$0.1~K.

The echo-detected field-swept EPR spectra were recorded at 85~K and 0~dB (300~W) microwave. The pulse sequence was $\pi/2$--$\tau$--$\pi$--$\tau$--echo, where $\tau$ and $\pi/2$-pulse are fixed at 200~ns and 12~ns, respectively. The delay $\tau$ and duration of $\pi/2$-pulse were also employed in the following experiments unless otherwise stated. The magnetic field scan step and scan range were set to be 2~Gauss and 1600~Gauss, respectively, corresponding to 801 points. The receiver gain was set to 54~dB. The same pulse sequence was used to monitor the dephasing time $T_2$ with 4~ns step and 1001 points. $T_1$ was measured with $\pi$--$t$--$\pi$/2--$\tau$--$\pi$--$\tau$--echo, where $t$ was initially 1000~ns and increased by 20~ns in each step.

Rabi oscillation measurements were conducted with pulse sequence of $t_p$--$t_0$--$\pi/2$--$\tau$--$\pi$--$\tau$--echo, where $t_0$ was fixed to be 600~ns, and $t_p$ was initially 2~ns and gradually increased with 2~ns in each step by totally 128 points. The echo intensity was integrated with gate length of 160~ns with 500 sampling averages. Rabi oscillation measurements were conducted at various microwave power, and for each power, the $\pi/2$ pulses were individually optimized.

\textbf{Calculations}

The first-principles calculations based on density functional theory (DFT) were performed with Vienna $Ab$ $initio$ Simulation Package (VASP) \cite{10.1103/PhysRevB.54.11169,20_PhysRevB.59.1758}. The Perdew-Burke-Ernzerhof (PBE) generalized gradient approximation (GGA) was used to describe the exchange-correlation interaction \cite{21_PhysRevB.50.17953,22_PhysRevLett.77.3865}. The density of $\Gamma$-centered k-point mesh is 0.3~\AA$^{-1}$. Cutoff energy of 520~eV and residual atomic forces criteria of 0.02~eV/\AA~ were selected in all simulations. The strong correlations of Cu $3d$ orbitals are included by the DFT+U method with $U_{\rm eff}=6.3$~eV \cite{PhysRevB.44.943,PhysRevB.57.1505}.

\section{Acknowledgments}

The authors gratefully acknowledge support from the National Natural Science Foundation of China (Grant Nos.~11974118, 12374107, 61975057, 21788102 and 51527804), the National Key R\&D Program of China (2020YFA0714604), the Foundation of Guangdong Province (2019B121205002), and the Key Research and Development Project of Guangdong Province (Grant No.~2020B0303300001).

\section{Author contributions}

X.Q. and Y.Y. designed and conducted the EPR experiments and analyzed the data with the help of C.H. in SCUT. J.L. synthesized and characterized the CSLA samples with the help of Z.Z. and J.W. in CSU. Y.P., G.C. and Y.Y. performed the simulations and theoretical analysis. Y.Y. wrote the manuscript with the help of other authors. X.B.Y. and D.M. contributed useful inputs to the work.

\section{Competing interests}

The authors declare no competing interests.

\bibliography{Echo_v6.bbl}

\begin{figure*}[h]
    \centering
    \includegraphics[width=1\linewidth]{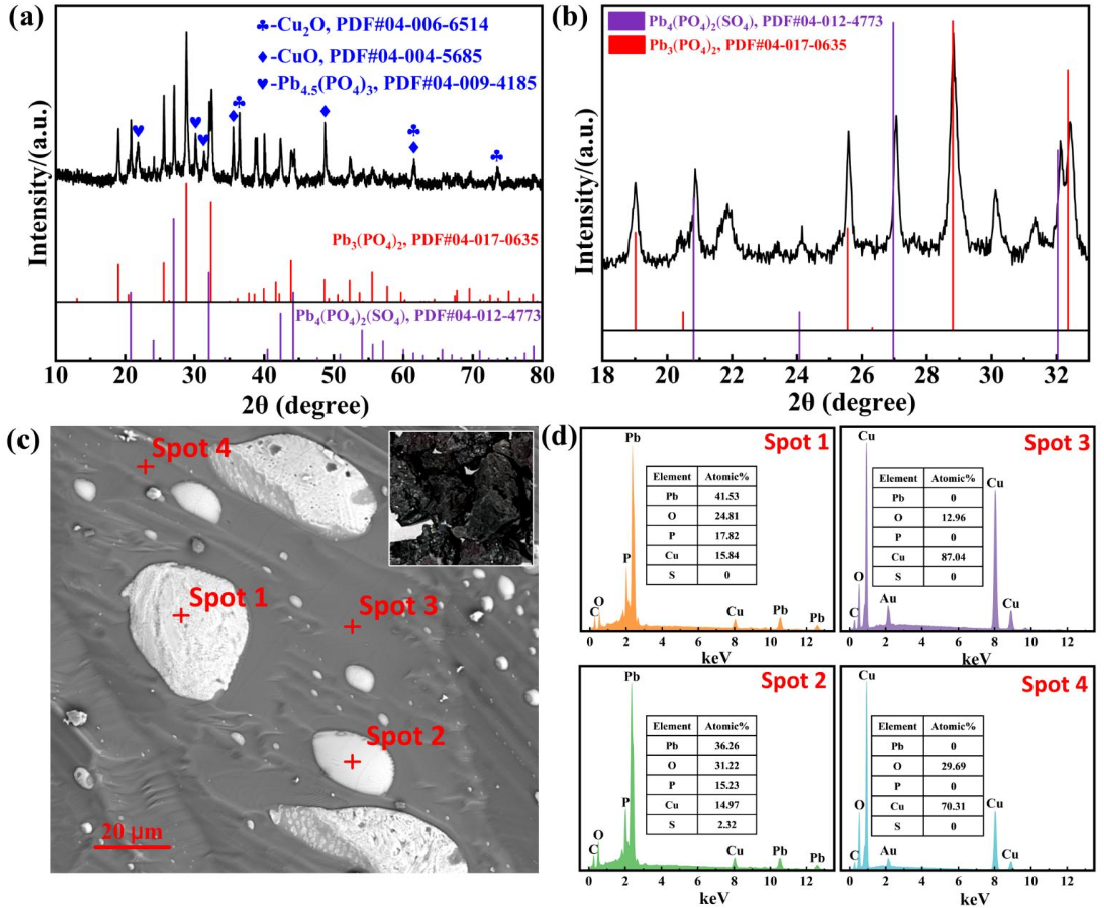}
    \caption{\textbf{Synthesis and structure.} (a) Powder XRD patterns of CSLA sample and the data of Pb3(PO4)2 and Pb4(PO)2(SO4) in Crystallography Database. (b) Magnified patterns at 2$\theta$ within 18--33$^{\rm o}$. (c) Scanning electron micrograph (SEM) at 2000X and the external photo (inset). (d) EDS patterns and atomic percentage of Spot 1 to 4 marked in SEM image.}
    \label{fig1}
\end{figure*}

\begin{figure*}[h]
    \centering
    \includegraphics[width=1\linewidth]{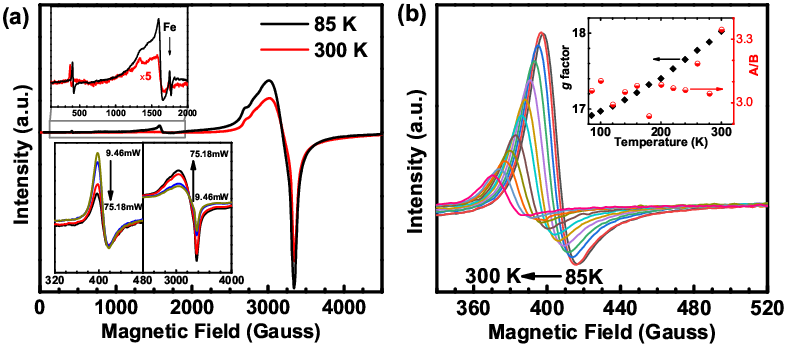}
    \caption{\textbf{cw EPR spectroscopy.} (a) EPR spectra at 85 and 300~K, with four remarkable signals: A wide peak from 2700 to 3400~Gauss, a sharp radical peak at 3350~Gauss, a half-field peak at 1620~Gauss and a low-field peak at 406~Gauss. Insets display the amplified peaks at high and low field, respectively. With increasing power of microwave from 9.46 to 75.18~mW, the signal normally increases at high field but strangely decreases and changes from asymmetric to symmetric at low field. The small peak at 1750~Gauss is from iron in quartz tube. (b) Low-field peaks at temperature of 85--300K. With increasing temperature, the peak position moves to lower field so that the relevant $g$ factor of the peak linearly increases, but the asymmetry ratio $A/B$ merely fluctuates in the error range (inset).}
    \label{fig2}
\end{figure*}

\begin{figure*}[h]
    \centering
    \includegraphics[width=1\linewidth]{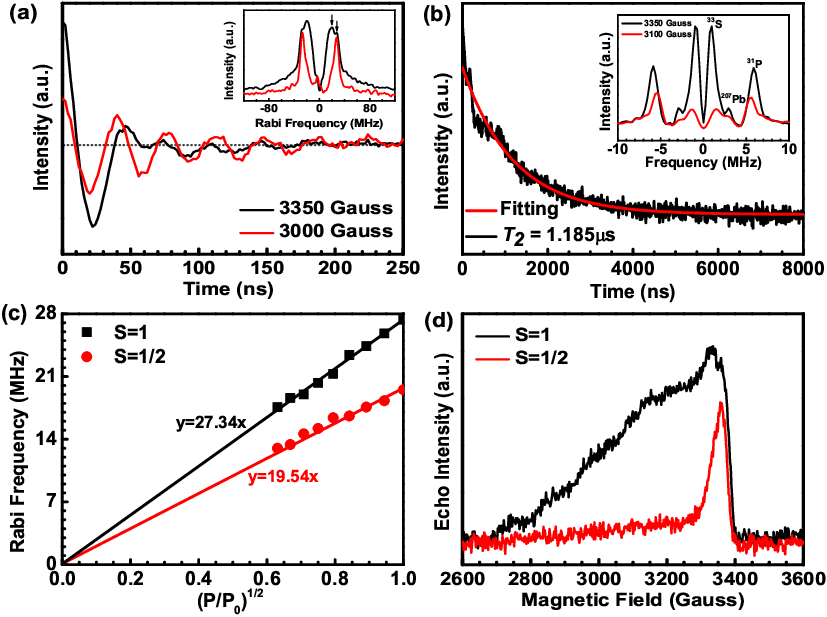}
    \caption{\textbf{Pulsed EPR spectroscopy.} (a) Rabi oscillations of the first seven periods at 3350 and 3000~Gauss and the relevant Fourier transformations (inset). As indicated by arrows, at 3350~Gauss there are two modes mixing with each other while at 3000~Gauss only one mode is present. (b) The coherence time $T_2$ is determined to be 1.185~$\mu$s by an exponential fitting. By Fourier transformations, inset shows the source of dephasing can be recognized as the neighboring sulphur, lead and phosphor. The copper and oxygen are absent suggesting the radicals are residing on cuprates and oxygen vacancies. (c) Rabi frequences of the two modes versus root of microwave power $P$, with $P_0$ being the power at 0~dB. The slope of the higher mode is closely $\sqrt{2}$ times larger than that of the lower mode indicating they can be assigned to spin one and half, respectively. (d) By changing the pulse duration of echo detection, the echo-detected field-swept spectra of spin one and half are respectively measured, which correspond to the wide and sharp radical peaks in cw EPR spectra. All curves were measured at 85~K.}
    \label{fig3}
\end{figure*}

\begin{figure*}[h]
    \centering
    \includegraphics[width=1.0\linewidth]{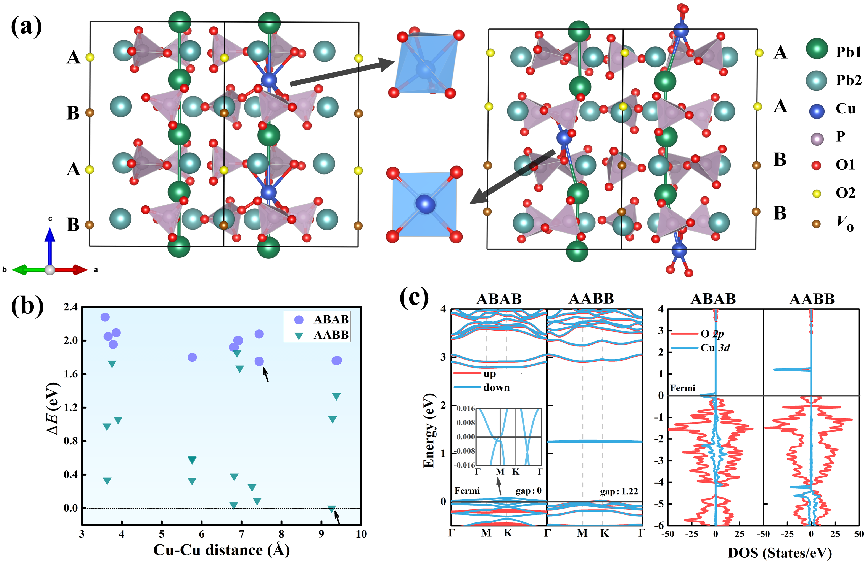}
    \caption{\textbf{First-principle simulations.} (a) Two stable structures and local Cu-O unit in copper-substituted ABAB and AABB. Labels A and B respectively represent O atoms and V$_{Os}$. There are two non-equivalent Pb atoms and two non-equivalent O atoms, labeled as Pb1, Pb2, O1, and O2. (b) The energy variation versus Cu-Cu distance in both ABAB and AABB structures. The $\Delta$$E$ is defined as the energy difference between the total energy of current configuration and the stablest one. Arrows indicate the most stablest structures. (c) The band structure and $p$DOS for the two stable structures calculated by DFT+U method. Inset shows the band structure close to the Fermi level in ABAB where a possible level crossing appears.}
    \label{fig4}
\end{figure*}

\end{document}